\documentclass[prl, twocolumn, superscriptaddress]{revtex4}

\usepackage{graphicx}
\usepackage{bm}
\newlength{\figwidth}
\usepackage{graphicx}
\usepackage{tabularx}
\usepackage{color}
\usepackage{amsmath}
\usepackage{array,multirow}
\usepackage{epstopdf}
\usepackage[author={Feng Guo}]{pdfcomment}
\usepackage[rgb]{xcolor}
\usepackage{xcolor}
\hypersetup{
    colorlinks,
    linkcolor={blue!80!black},
    citecolor={blue!80!black},
    urlcolor={blue!80!black}
}

\begin{document}
\title{Ferromagnetic resonance phase imaging in spin Hall multilayers}

\author{Feng Guo}
\affiliation{School of Applied and Engineering Physics, Cornell University, Ithaca, NY 14853, USA}

\author{Jason M. Bartell}
\affiliation{School of Applied and Engineering Physics, Cornell University, Ithaca, NY 14853, USA}

\author{Gregory D. Fuchs}
\email{gdf9@cornell.edu}
\affiliation{School of Applied and Engineering Physics, Cornell University, Ithaca, NY 14853, USA}

\begin{abstract}

We experimentally image the magnetic precession phase of patterned spin Hall multilayer samples to study the rf driving field vector using time-resolved anomalous Nernst effect (TRANE) microscopy. Our ferromagnetic resonance (FMR) measurements quantify the phase and amplitude for both the magnetic precession and the electric current, which allows us to establish the total driving field orientation and the strength of spin Hall effect. In a channel of uniform width, we observe spatial variation of the FMR phase laterally across the channel. We interpret our findings in the context of electrical measurement using the spin-transfer torque ferromagnetic resonance technique and show that observed phase variation introduces a systematic correction into the spin Hall angle if spatial phase and amplitude variations are not taken into account.

\end{abstract}

\maketitle



 
When a spin current traverses the interface between a normal metal and a ferromagnetic metal, it generates a spin-transfer torque\cite{Slonczewski_STT_JMMM96, Berger_STT_prb96, RalphS_STT_JMMM08} that efficiently manipulates magnetization. Accurate quantification of current-induced torques is pivotal to first understanding and then engineering spintronic devices for future magnetic memory and information technology.  Experimentally, most studies of spin Hall effect (SHE)-driven torques\cite{DyakonovP_SHE_PLA71, Hirsch_SHE_prl99, KatoMGA_SHE_science04} have relied on electrical measurements of devices, which are effective because they provide a large signal-to-noise ratio.  Typical electrical techniques are spin torque ferromagnetic resonance (ST-FMR)\cite{TulapurkarSFKMTDWY_ST-FMR_nature05, SankeyBGKBR_ST-FMR_prl06, LiuTRB_ST-FMR_prl11} for in-plane magnetic moments, and harmonic Hall voltage analysis\cite{PiKBLCLSS_apl10_2ndHarmonics, KimSHYFSMO_NatMat13_2ndHarmonics, PaiNBVRB_apl14_2ndHarmonics, HayashiKYO_prb14_2ndHarmonics} for perpendicularly magnetized devices.  An essential assumption of these methods is that both the driving current and the magnetic response are uniform. To gain deeper understanding of SHE-driven torques and go beyond approximate treatments, we quantify the relationship between the driving current and the dynamic magnetic response using phase sensitive magnetic microscopy. Our measurements show that while the assumption of uniform driving current is valid, the assumption of uniform magnetic response is not.
 
Dynamic magnetic imaging provides a method of verifying the uniformity of a magnetic response and measuring spin torques.  Several techniques have been developed  to sense local magnetization dynamics in micro- and nano-structures, including microfocused Brillouin light scattering (BLS)\cite{PerzlmaierBBDHD_BLS_Kerr_prl05, DemokritovD_BLS_Transmag08, DemidovBRKMRD_BLSnonlinear_prl10, NembachSSJKMK_BLS_prb11}, ferromagnetic resonance force microscopy (FMRFM) dynamics imaging\cite{KleinDNBGHLSTV_fmrfm_prb08, LeeOXHYBPH_fmrfm_nature10, ChiaGBM_localization_prl12, GuoBM_edge_prl13}, time-resolved magneto-optical Kerr effect (MOKE) microscopy\cite{HiebertSF_kerr_prl97, ParkEEBC_kerr_prl02}, and x-ray magnetic circular dichroism (XMCD)\cite{ChoeASBDSP_xmcdVortex_science04, MarchamKNHCSVTCKSA_xray_jap11, WesselsEWNVVMWD_xmcdDomains_prb14}.  Additionally, an optical technique based on polar MOKE for measuring a dc-driven spin-torque vector was recently introduced\cite{FanCWNLX_NatComm14_DelawareMOKE, FanMWRWCLRX_DelawareCornellQuadraticMOKE}.  However, very few phase-sensitive imaging techniques provide a full set of information --- a quantitative image of both drive and magnetic response up to gigahertz frequencies.


In this work, we use time-resolved anomalous Nernst effect (TRANE) microscopy\cite{BartellDLF_NatComm15_TRANE, GuoBNF_prapplied15_TRANEtechnique} to simultaneously image ferromagnetic resonance (FMR) and rf driving current in spin Hall multilayers.  By imaging the amplitude and phase of the precessing magnetization in relation to the driving current, we find that the driving field direction in a sample with strong spin torque is different than in a sample where the spin torque is blocked with a 2 nm thick Hf spacer. More importantly, we demonstrate that even in a uniform width structure, the FMR phase is nonuniform, despite the common assumption of quasi-uniformity.  We analyze the consequences of spatial variations in precession phase in terms of device-level measurements such as ST-FMR.  We show that ST-FMR measurements of the spin Hall efficiency can have a sizable systematic error, depending on the details of the sample.


 \begin{figure}[tb]
  \centering
  \graphicspath{{../Figures/}}
  \includegraphics[width=\figwidth]{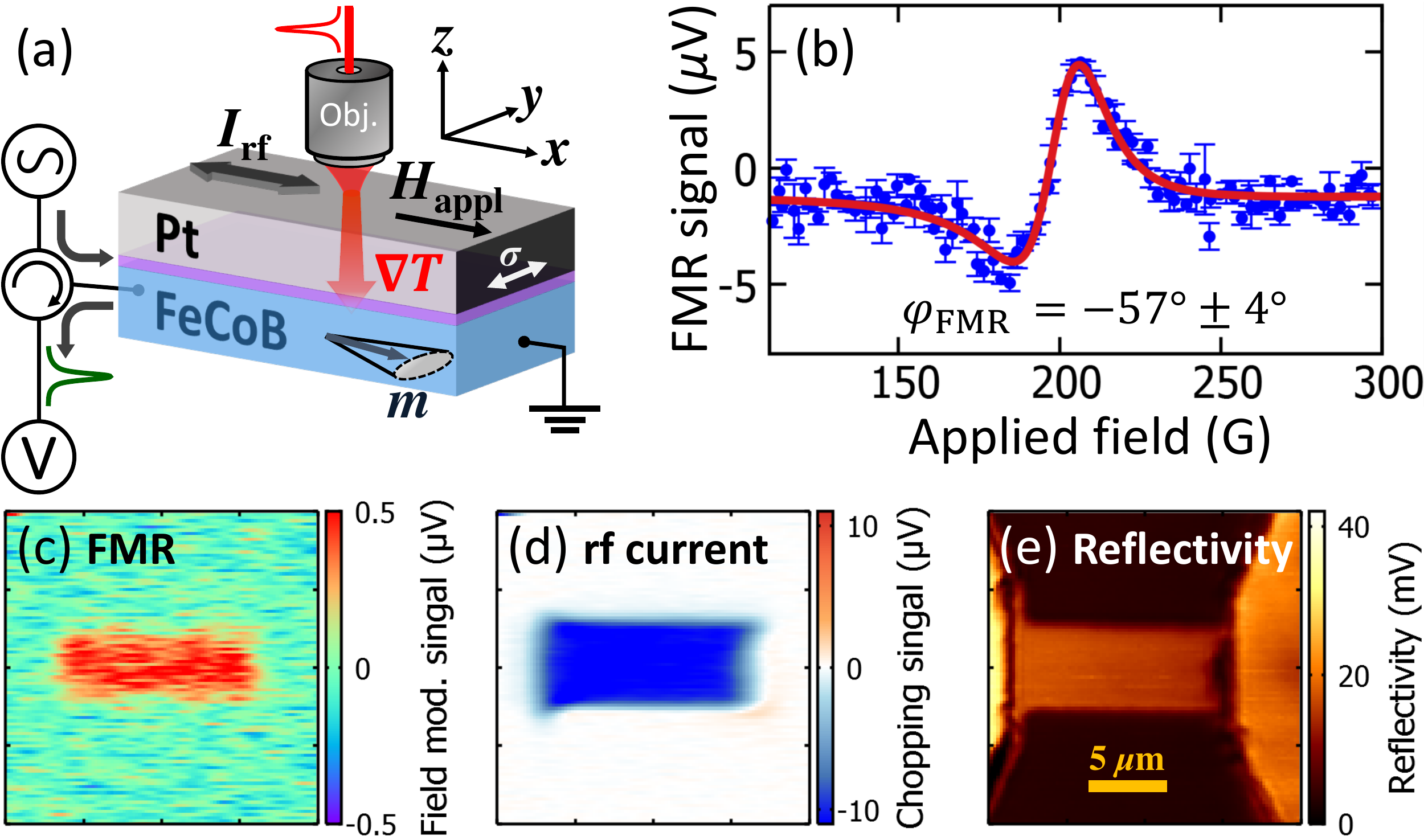}
  \caption{{\bf  Time-resolved anomalous Nernst effect (TRANE) microscopy measurement concept and examples of spectroscopical and imaging measurements.} (a) Schematics of the measurement principle for TRANE microscopy. (b) Example of a ferromagnetic resonance (FMR) spectrum using field modulation. Images measured at a fixed field of 208~G: (c) field modulated FMR signal, (d) rf current with chopping signal and (e) laser reflectivity.
    \label{fig:setup}}
\end{figure}

To simultaneously probe the local magnetic orientation and the rf driving current, we use 3~ps long pulsed laser heating, as illustrated in Fig.~\ref{fig:setup}(a). The magnetization projection in $y$ direction is detected through the anomalous Nernst effect.  An increase in the local resistivity due to the transient heating also produces a voltage corresponding to the driving current. Fig.~\ref{fig:setup} (b) shows an example of an FMR spectrum, from which we obtain the phase of FMR precession. With a fixed magnetic field, we can simultaneously image the FMR signal, the rf current signal, and the laser reflectivity (Figs.~\ref{fig:setup} (c) - (e), respectively). For this study, we fabricate samples with a stack structure of ${\rm Fe_{60}Co_{20}B_{20}(4~nm)/Hf (t_{\rm Hf})/Pt(4~nm)}$. We use two hafnium thicknesses, ${\rm t_{Hf}}$=0.3~nm and ${\rm t_{Hf}}$=2~nm. The 0.3~nm Hf samples, which we will simply refer to as the ``spin Hall samples'', present a reasonably large spin Hall effect while maintain a low damping parameter. In contrast, the samples with 2~nm thick Hf, or the ``non spin Hall samples'', have a minimal spin Hall effect. The spin Hall efficiencies of these two sets of samples are also confirmed with ST-FMR measurements, summarized in Table~\ref{tab:comparison}. As discussed later, we use the non spin Hall sample to establish the local driving field angle in the spin Hall sample via precesion phase measurements.



First we analyze the effective driving field angle with respect to the sample plane, $\theta_{\rm eff}$, from measurements of the FMR precession phase. Specifically,  $\theta_{\rm eff}$, is directly related to the  difference between the FMR phase and the driving current phase according to the relation $\theta_{\rm eff}=\varphi_{\textsc{\tiny FMR}}-\varphi_{\rm rf}+90^\circ$. In general, the FMR phase can be written as:
\begin{subequations}
\label{eq:FMRphase}
\begin{align}
         \varphi_{\textsc{\tiny FMR}}^\pm&=\pm (\varphi_{\rm rf}+\theta^\pm) -90^\circ,         
         \\
      \text{Intersection: }&
      \begin{cases}
		\varphi_{\rm rf}^{\rm int}= -(\theta^+ + \theta^-)/2, \\
		\varphi_{\textsc{\tiny FMR}}^{\rm int}= (\theta^+ - \theta^-)/2 - 90^\circ,
	\end{cases}		
\end{align}
\end{subequations}
in which the FMR phase $\varphi_{\textsc{\tiny FMR}}$ simply follows the current phase $\varphi_{\rm rf}$ and the driving field angle $\theta$. The superscripts ``$+$'' and ``$-$'' denote the positive and negative field directions respectively. 



\begin{figure}[tb]
  \centering
  \graphicspath{{../Figures/}}
  \includegraphics[width=\figwidth]{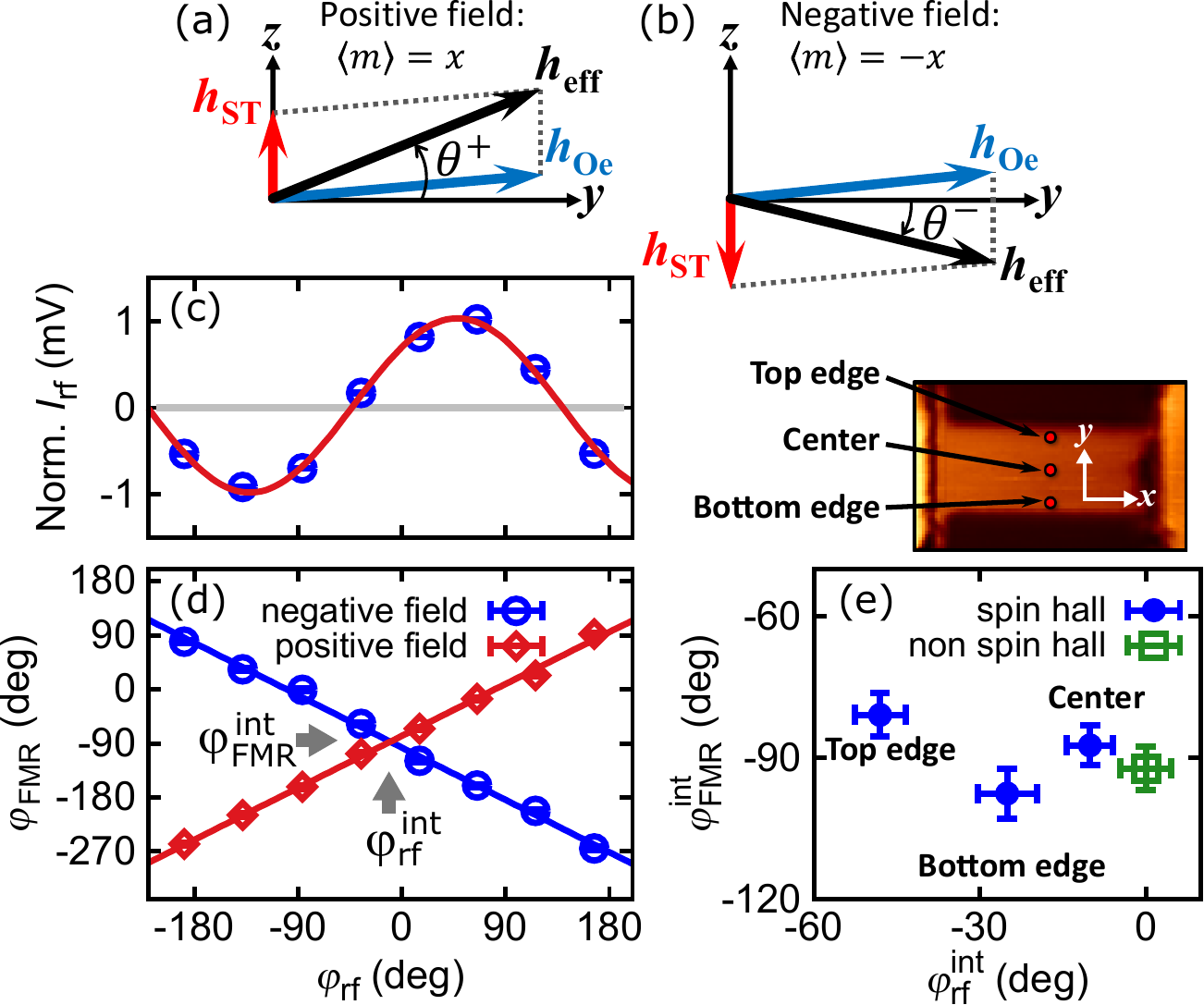}
  \caption{{\bf Vector diagrams and FMR phase measurements.} Diagrams of spin torque field $h_{\rm ST}$ and Oersted field $h_{\rm Oe}$ for (a) positive and (b) negative applied fields. The charge current is set to the positive ($+\hat{x}$) for both cases. Normalized current signal (c) and FMR phase (d) as functions of rf current phase. Both (c) and (d) are  measured at the center the of bar.  (e) The points of intersection measured at the top edge, center and bottom edge of the channel. The intersection measured at the center of a non spin Hall sample is also included (hollow square) in (e). 
    \label{fig:FMRphaseCenter}}
\end{figure}

To discuss the physical meanings of the intersection ($\varphi_{\rm rf}^{\rm int}$, $\varphi_{\textsc{\tiny FMR}}^{\rm int}$), we first explain the symmetries of various torques. Let us consider a case where the total driving torque has both Oersted and spin torque (anti-damping like) contributions\footnote{We do not include a field-like spin torque in the discussion since it is indistinguishable from the in-plane Oersted field using phase measurements only.}. As illustrated in Figs.~\ref{fig:FMRphaseCenter} (a) and (b), the Oersted field $\hat{h}_{\rm Oe}$ does not change sign when the field reverses, while the spin torque driving field $\hat{h}_{\rm ST}=\hat{m}\times\hat{\sigma}$ does.

Due to the difference in the symmetry of $\hat{h}_{\rm Oe}$ and $\hat{h}_{\rm ST}$, the two coordinates at the intersection in Eq.~\ref{eq:FMRphase}b have different physical interpretations.  $\varphi_{\rm rf}^{\rm int}$ is the averaged effective field angle.  $\varphi_{\textsc{\tiny FMR}}^{\rm int}$ is determined by the difference between $\theta^+$ and $\theta^-$, which is sensitive to the Oersted field orientation. In the Supplementary Information\cite{Supplemental} we further show that  $\varphi_{\rm rf}^{\rm int} \approx -{\textstyle h_{\rm ST}}/{\textstyle h_{\rm Oe}^\parallel}$ and  $\varphi_{\textsc{\tiny FMR}}^{\rm int} \approx  {\textstyle h_{\rm Oe}^\perp / {\textstyle h_{\rm Oe}^\parallel}} - 90^\circ$, under the assumption of $h_{\rm ST}/h_{\rm Oe}^\parallel$, $h_{\rm Oe}^\perp/h_{\rm Oe}^\parallel \ll 1$ ($h_{\rm Oe}^\perp$ and $h_{\rm Oe}^\parallel$ are the out-of-plane and in-plane components of the Oersted field, respectively).


Figs.~\ref{fig:FMRphaseCenter} (c) and (d) show the phase dependent rf current and FMR signals, measured at the center of the channel. The rf current signal shows a sinusoidal waveform. The FMR phase varies linearly with increasing rf current phase, with a slope of $+1$ ($-1$) for the positive (negative) applied field, consistent with Eq.~\ref{eq:FMRphase}a. To investigate the spatial dependence of the phase, we repeat the measurement in Fig.~\ref{fig:FMRphaseCenter} (d) for the top and bottom edges of the channel. The points of intersection $(\varphi_{\rm rf}^{\rm int}, \varphi_{\textsc{\tiny FMR}}^{\rm int})$ for different positions in the spin Hall sample are plotted in Fig.~\ref{fig:FMRphaseCenter} (e). We also include the intersection measured at the center of the non spin Hall sample (2~nm Hf spacer) as reference. In the absence of the spin Hall effect, only Oersted field is responsible for the effective driving field, and we expect it to be nearly in-plane at the center of the channel. When the spin torque is turned on, given the stack sequence and the positive spin Hall angle for platinum, we expect $\hat{h}_{\rm ST}\parallel+\hat{z}$. As a result, the spin torque tilts $\theta_{\rm eff}$ out of the sample plane, towards the $+\hat{z}$ direction. As the driving field angle increases, $\varphi_{\rm rf}^{\rm int}$ will decrease, in agreement with Fig.~\ref{fig:FMRphaseCenter} (e).

Because the FMR phase and current phase do not share an absolute reference\cite{Supplemental}, we use the non spin Hall sample to define the zero current phase, by assuming that the non spin Hall sample has an in-plane driving field at the channel center [i.e. $\varphi_{\rm rf}^{\rm int}=0$, see the green point in Fig.~\ref{fig:FMRphaseCenter} (e)]. We also assume that the temporal profiles of both the temperature and thermal gradient remain the same between the two samples, since they are nearly identical structures\footnote{We do not expect the extra 1.3~nm Hf in the non spin Hall sample to significantly change the temperature and thermal gradient relaxation timescales.}. Finally using the FMR phase of the non spin Hall sample, we obtain a driving field angle of $\theta_{\rm eff}^0=10.1^\circ\pm4.2^\circ$ for the spin Hall sample, corresponding to a $(J_s/J_c)^0=0.048\pm0.020$. The comparisons between ST-FMR electrical measurements and FMR phase measurements are shown in Table~\ref{tab:comparison}.

\begin{figure}[tb]
  \centering
  \graphicspath{{../Figures/}}
  \includegraphics[width=\figwidth]{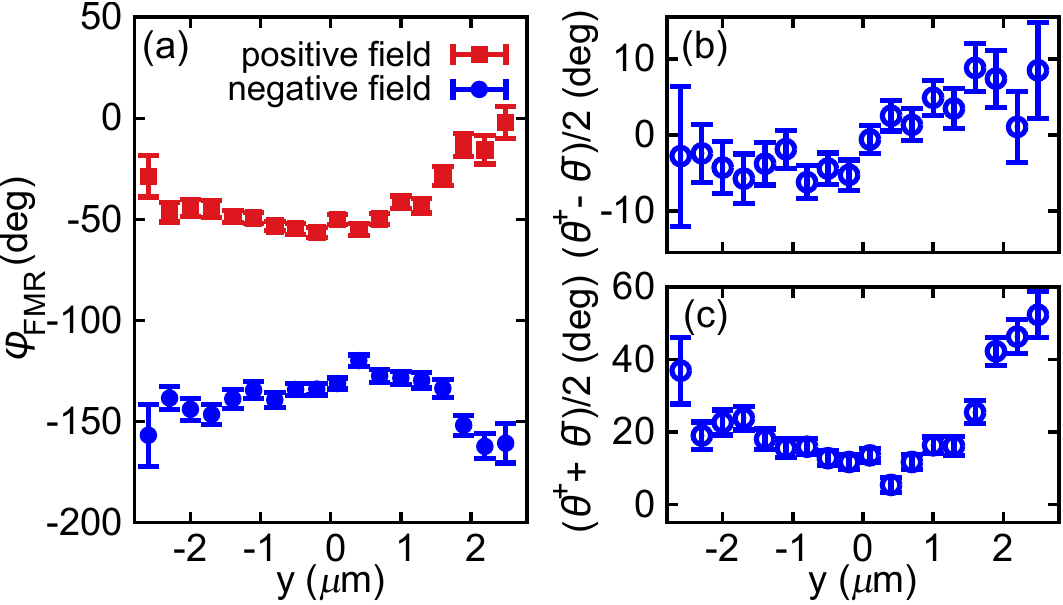}
  \caption{{\bf A significant spatial variation in the FMR phase along $y$ direction.}  (a) FMR phases measured at positive (red) and negative (blue) applied fields as functions of $y$ position. Using Eq.~\ref{eq:FMRphase}, we can also plot (b) $(\theta^+ - \theta^-)/2$ and (c) $(\theta^+ + \theta^-)/2$ as functions of $y$. 
    \label{fig:FMRphasevsy}}
\end{figure}

In the following, we focus on the position dependent FMR phase for the spin Hall sample. As shown in Fig.~\ref{fig:FMRphaseCenter} (e), $\varphi_{\rm rf}^{\rm int}$ near either the top or bottom edge is less than that measured at the center, indicating a larger $\theta_{\rm eff}$ at the edges. 
However $\varphi_{\textsc{\tiny FMR}}^{\rm int}$ near either the top or bottom edge shifts towards opposite directions with respect to the center, suggesting a gradual change in $h_{\rm Oe}^\perp/h_{\rm Oe}^\parallel$. Furthermore, $h_{\rm Oe}^\perp$ is expected to be positive at the top edge and negative at the bottom edge, which is consistent with the vertical sequence of the three points in  Fig.~\ref{fig:FMRphaseCenter} (e).


To further investigate phase variation across the width of the channel, we measure the FMR spectrum as a function of $y$ position for both positive and negative applied fields, with a fixed rf current phase. The $y$ dependent FMR phase is shown in Fig.~\ref{fig:FMRphasevsy}. There is a sizable variation of $\sim50^\circ$ along the $y$ direction. We use Eq.~\ref{eq:FMRphase} to plot the difference and sum of $\theta^+$ and $\theta^-$, as functions of $y$. Consistent with the previous measurements in Fig.~\ref{fig:FMRphaseCenter} (e), $(\theta^+ + \theta^-)/2$ is larger near the sample edges. While $(\theta^+- \theta^-)/2$ is also in agreement with the vertical positions of the points in  Fig.~\ref{fig:FMRphaseCenter} (e).

\begin{figure}[tb]
  \centering
  \graphicspath{{../Figures/}}
  \includegraphics[width=\figwidth]{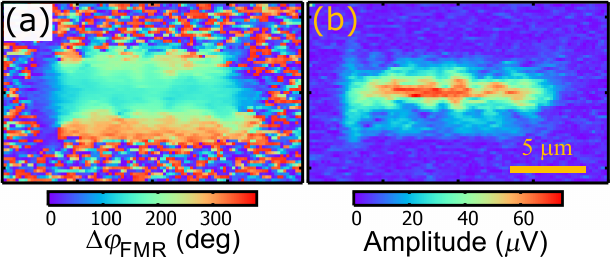}
  \caption{{\bf Images of the FMR phase and amplitude.} By fixing $\varphi_{\textsc{\tiny FMR}}=-24^\circ$ and using 6 FMR images at various applied fields, we can decompose (a) the relative FMR phase variation and (b) the FMR amplitude.
    \label{fig:FMRimaging}}
\end{figure}

In addition, we demonstrate an approach to image the FMR phase. Instead of recording the FMR spectrum at each location, we combine multiple FMR images to calculate the phase variation. In this example, we fix the rf current phase corresponding to a FMR phase of $-24^\circ$ at the channel sample. We then combine 6 FMR images at various applied fields (from 185~G to 215~G), to reconstruct both phase and amplitude images, shown in Fig.~\ref{fig:FMRimaging} (a) and (b) respectively. The main feature of the phase image is that the phase is quasi-uniform near the center, and it increases near the edges, in quantitative agreement with data in Fig.~\ref{fig:FMRphasevsy}. The phase variation is more prominent along the $y$-direction than that in the $x$-direction. In contrast, the FMR amplitude is large near the center and decreases towards either edge, as expected.

\begin{figure}[tb]
  \centering
  \graphicspath{{../Figures/}}
  \includegraphics[width=\figwidth]{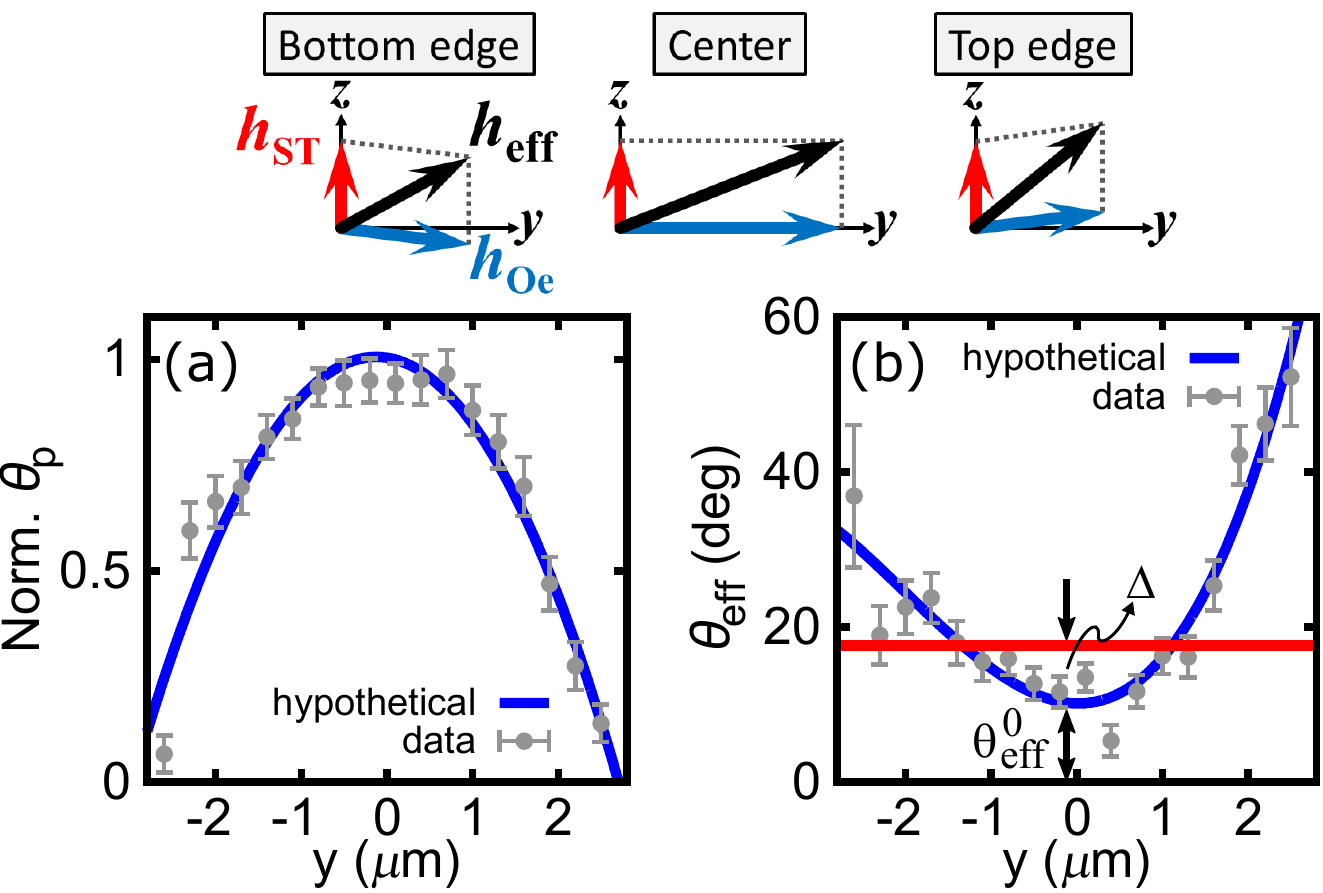}
  \caption{{\bf Numerical estimation of the phase correction given the spatial profiles of precession amplitude and phase.} Spatial distributions of (a) the normalized precession amplitude and (b) effective driving field angle. The blue curves in (a) and (b) are the polynomial fits of the data (gray points). The red line in (b) is the resultant driving field angle that would be obtained from ST-FMR, using Eq.~\ref{eq:st-fmr_phase}.
    \label{fig:st-fmr_correction}}
\end{figure}

Next we speculate on the origin of the phase variation, then we evaluate its influence on global electrical measurements. In a confined magnetic structure under a uniform applied field, the internal magnetic field is highly nonuniform near the edges due to  the demagnetizing field. The effects of nonuniform internal field and hence nonuniform precession modes are well established for magnetic micro- and nanostructures\cite{JorzickDHBFGSBG_internalfield_prl02, BayerDHS_internalfield_apl03, BayerPWYCC_internalfield_prb04, McMichaelM_edgeInternalfield_prb06, NembachSBS_internalfield_prl13}. Similarly, for the bar structure samples used in this study, the transverse driving field experiences an inhomogeneous demagnetizing field. Consequently the in-plane driving field $h_{\rm Oe}^\parallel$ inside the ferromagnet\footnote{All the transverse in-plane driving fields in the system will experience such inhomogeneous demagnetizing field, for both in-plane Oersted field ($h_{\rm Oe}^\parallel$) and possible field like contribution ($h_{\rm FL}$).} has a substantial spatial variation, which plays an important role in the observed phase variation. In contrast, the out-of-plane $h_{\rm ST}$ is uniform across the sample given a uniform current density distribution. As a result, $h_{\rm Oe}^\parallel$ is weaker at the channel edges and the effective driving field close to the edges has a larger angle than that at the center, as illustrated in the inset of Fig.~\ref{fig:st-fmr_correction}. We argue that the spatial variation of the rf driving field is determined by a number of factors, including sample dimensions, rf current uniformity, edge properties, and magnetic anisotropy fields. Therefore the details of the phase variations are expected to be sample specific.


Now we examine how a nonunifrom precession phase can affect the result characterized by ST-FMR. The analysis of ST-FMR relies on two assumptions: one is the uniformity of the precession mode and the other is the uniformity of the rf current.  Although the measured driving current is uniform, the assumption of uniform precession breaks down. We start with the signal measured by ST-FMR under the macrospin spin approximation. By mixing the rf current with an oscillating magnetoresistence, we get a rectified voltage: $ V_{\rm mix} \propto \theta_{\rm p} \left \{
\chi'(H) \cos (\varphi_{\rm rf} - \varphi_{\textsc{\tiny FMR}})+ \chi''(H) \sin (\varphi_{\rm rf} - \varphi_{\textsc{\tiny FMR}}) 
\right \}$, where $\theta_{\rm p}$ is the precession amplitude and $\chi'$ and $\chi''$ are the real and imaginary susceptibility functions respectively.  By fitting the spectrum $V_{\rm mix}(H)$ to a linear combination of the symmetric and anti-symmetric Lorentzian functions, one obtains $\varphi_{\rm rf} - \varphi_{\textsc{\tiny FMR}}$ and thus the spin Hall efficiency for the normal metal/ferromagnet combination.

To include the effect of spatial variation in $\varphi_{\textsc{\tiny FMR}}$, we rewrite the averaged $V_{\rm mix}$ as the integral of the mixing voltage weighted by the precession amplitude ($\theta_{\rm p}$) (see Supplementary Information\cite{Supplemental} for derivation):
\begin{align}
  \label{eq:Vmix}
V_{\rm mix} & \propto \chi' \int {\rm d}r~\theta_{\rm p}(r)  \cos \big[ \varphi_{\rm rf} - \varphi_{\textsc{\tiny FMR}}(r) \big]\nonumber \\
  &+ \chi'' \int {\rm d}r~\theta_{\rm p}(r)  \sin \big[ \varphi_{\rm rf} - \varphi_{\textsc{\tiny FMR}}(r) \big].
\end{align}
Therefore, we find the equivalent phase difference between the FMR and rf current that would be obtained from the global measurement is:
\begin{align}
  \label{eq:st-fmr_phase}
&\langle \varphi_{\textsc{\tiny FMR}}-\varphi_{\rm rf}\rangle=\nonumber\\
&~~~~~~\tan^{-1}\left[\frac{\int {\rm d}r~\theta_{\rm p}(r)  \sin\big[ \varphi_{\textsc{\tiny FMR}}(r)- \varphi_{\rm rf} \big]}{\int {\rm d}r~\theta_{\rm p}(r)  \cos\big[ \varphi_{\textsc{\tiny FMR}}(r)- \varphi_{\rm rf} \big]}\right] 
\end{align}

As an interesting point, the phase correction $\Delta$ is only determined by the spatial varying component of the FMR phase $\delta\varphi_{\textsc{\tiny FMR}}(r)$ and the precession amplitude $\theta_{\rm p}(r)$, and is independent of the overall offset $\varphi_{\textsc{\tiny FMR}}^0$.  We can substitute  $\varphi_{\textsc{\tiny FMR}}^0+\delta\varphi_{\textsc{\tiny FMR}}(r)$ for $\varphi_{\textsc{\tiny FMR}}$ in Eq.~\ref{eq:st-fmr_phase}. Using trigonometry, we can rewrite Eq.~\ref{eq:st-fmr_phase}:
\begin{align}
  \label{eq:st-fmr_phase2}
	&\langle \varphi_{\textsc{\tiny FMR}}-\varphi_{\rm rf}\rangle =\varphi_{\textsc{\tiny FMR}}^0- \varphi_{\rm rf}+\Delta, \\
	&\Delta=\tan^{-1}\left[ \frac{\int {\rm d}r~\theta_{\rm p}(r)  \sin \big[ \delta\varphi_{\textsc{\tiny FMR}}(r) \big]}{\int {\rm d}r~\theta_{\rm p}(r)\cos  \big[  \delta\varphi_{\textsc{\tiny FMR}}(r) \big]}\right]. \nonumber
\end{align}
Therefore the correction for the phase (also driving field angle) $\Delta$ is independent of the driving field angle near the channel center $\theta_{\rm eff}^0$.

To numerically evaluate the correction resulting from the phase nonuniformity, we use the polynomial fits of the precession amplitude $\theta_{\rm p}(y)$ and driving field angle $\theta_{\rm eff}(y)$ to mimic the experimental results, plotted in Fig.~\ref{fig:st-fmr_correction}. For simplicity we assume both the phase and amplitude of the precession are uniform in the $x$ direction and we only consider the spatial variation along the $y$ direction. We get an ``averaged'' value of the effective driving field, $\langle \theta_{\rm eff} \rangle=\theta_{\rm eff}^0+\Delta$, shown as the red line in Fig.~\ref{fig:st-fmr_correction} (b). In this example, the phase correction $\Delta= 7.5^\circ\pm 1.8^\circ$.

As shown in Table~\ref{tab:comparison}, for the spin Hall sample there is a substantial discrepancy between  $(J_s/J_c)^0$ at the channel center and the ST-FMR results. At the same time, $\langle J_s/J_c \rangle$ obtained by including the the spatially varying phase is consistent with $J_s/J_c$ measured from ST-FMR, within experimental errors. Thus, we argue that the ST-FMR technique does not necessarily reflect the phase value in the middle of the sample, nor does it sense a uniform phase, rather it provides a spatially averaged phase. Although the electrical techniques have a superior sensitivity, we show an example where it is essential to include a correction for spatial variations of both precession phase and amplitude to correctly quantify the spin Hall efficiency from electrical measurements.

\begin{table}[tb]
\centering
\caption{{\bf Comparison between electrical ST-FMR and spatially resolved measurements.} The spin Hall efficiency, $J_s/J_c$, measured with ST-FMR and calculated from the spatial variation of the FMR phase. $\theta_{\rm eff}^0$ is the angle of effective driving field at the center, and $\Delta$ is the phase correction due to the phase variation.  $(J_s/J_c)^0$ is the ratio only at the center, corresponding to $\theta_{\rm eff}^0$; while $\langle J_s/J_c \rangle$ is integrated ratio with the phase variation included. The uncertainty of $\Delta$ is calculated using the standard errors of the polynomial fits in Fig.~\ref{fig:st-fmr_correction}.}
\begin{tabular}{ l r c c}
\\[-8pt]
 \hline\hline
 \\[-8pt]
\multicolumn{2}{l}{\multirow{2}{*}{Sample}} & {\bf spin Hall} & {\bf non spin Hall} \\  
 	&& (0.3~nm Hf) &  (2~nm Hf) \\ \hline
\\[-6pt]
ST-FMR & $J_s/J_c=$ & 0.076$\pm$0.002 & 0.010$\pm$0.003 \\ \hline
\\[-6pt]
\multirow{4}{*}{\rotatebox[origin=c]{90}{\parbox{1.5cm}{Phase var. included}}} & $\theta_{\rm eff}^0=$ & $10.1^\circ \pm 4.2^\circ$ & assume: 0 \\ 
	& $\Delta=$ & $7.5^\circ \pm 1.8^\circ$ & $3.4^\circ \pm 1.9^\circ$ \\
	& $(J_s/J_c)^0=$ & $0.048\pm0.020$ & assume: 0 \\
	& $\langle J_s/J_c \rangle=$ & $0.086\pm0.031$ & $0.015\pm0.008$ \\[2pt]

\hline \hline 

\end{tabular}
\label{tab:comparison}
\end{table}


In summary, we have studied the FMR phase in uniform width spin hall multilayers. Using TRANE microscopy, we have measured the amplitude and phase of both FMR precession and rf driving current, which enables us to determine the angle of driving field vector. In a sample with substantial spin torque, we found that the driving field points around $10^\circ$ out of the sample plane at the center. More importantly, we observe a substantial precession phase variation across the width of the channel. We expect the phase variation to be important in the micro- and nano-structures, depending on device details. We have also evaluated the correction term in driving field angle $\Delta$ due to the phase variation. In the case of nonuniform precession phase, we have established a mechanism by which ST-FMR can yield an inaccurate spin Hall efficiency. For the 5~$\mu$m wide samples studied, when including the spatial variations of the precession amplitude and phase, the integrated spin Hall efficiency nearly doubles compared to that in the middle of the channel.  Therefore, although electrical measurements are very effective techniques to quantify the spin Hall effect, we conclude the spatial variations of both precession amplitude and phase can play an important role and should not be overlooked. Finally, we have shown that phase-sensitive imagining techniques such as TRANE microscopy are valuable for quantitative studies of the spin Hall effect. The spatial uniformity is also an essential ingredient for understanding damping and switching dynamics in magnetic confined structures.

\section*{Methods}

\subsection{\footnotesize{Basic principles of time-resolved anomalous Nernst effect (TRANE) microscopy}}
We apply an rf current through a circulator to the multilayer sample to excite magnetic dynamics, shown in Fig.~\ref{fig:setup} (a). We use pulsed (3~ps) laser heating, synchronized with the phase of the current source to enable stroboscopic measurements of both the magnetization projection $m_y$ and the rf driving current. The transient vertical thermal gradient produces a voltage pulse corresponding to the magnetization projection in $y$ direction, through the anomalous Nernst effect. The transient heating also increases the local resistivity, which produces a voltage when a gigahertz driving current is applied. An external field is applied along $\hat{x}$ direction. When measuring the FMR spectra, the in-plane magnetization component, $m_y$, is recorded as a function of applied field. We also apply a modulation field to distinguish the magnetic signal from the rf current signal. We establish the phase of FMR precession by fitting the signal to $A(\frac{{\rm d}\chi'(H)}{{\rm d}H} \cos\varphi_{\textsc{\tiny FMR}}+\frac{{\rm d}\chi''}{{\rm d}H}\sin\varphi_{\textsc{\tiny FMR}})$, where $\chi'$ and $\chi''$ are the real and imaginary dynamic susceptibility functions, $A$ is the local FMR amplitude, and $\varphi_{\textsc{\tiny FMR}}$ is the FMR phase at resonance. In the example shown in Fig.~\ref{fig:setup} (b), we find $\varphi_{\textsc{\tiny FMR}}=-57^\circ\pm4^\circ$. To image dynamics in this sample, we set the applied field to 208~G and record the FMR, the rf current, and laser reflectivity, as shown in Fig.~\ref{fig:setup} (c), (d) and (e) respectively. A more detailed description of the TRANE technique can be found in prior work\cite{GuoBNF_prapplied15_TRANEtechnique}. 

\subsection{\footnotesize{Sample fabrication and characterization}}
The samples were dc sputtered on the thermally conductive sapphire substrates and subsequently patterned into a ${\rm 5~\mu m \times 12~\mu m}$ bar geometry using photolithography. The multilayer samples have a stack structure of ${\rm Fe_{60}Co_{20}B_{20}(4~nm)/Hf (t_{\rm Hf})/Pt(4~nm)}$. Two hafnium thicknesses are used in this study, ${\rm t_{Hf}}$=0.3~nm and ${\rm t_{Hf}}$=2~nm. The 0.3~nm Hf samples (``spin Hall samples'') have a substantial spin Hall effect and a low damping parameter. From a previous study, a thin Hf spacer layer (near 0.5~nm) is helpful to enhance the spin Hall effect efficacy\cite{NguyenPNMRB_Hf_APL15}. The 2~nm thick Hf samples (``non spin Hall samples''), however, have a minimal spin Hall effect. Since the thickness of the Hf spacer already exceeds the spin diffusion length of 1.5~nm in Hf\cite{PaiNBVRB_apl14_2ndHarmonics}, the Hf layer blocks the spin current flowing from the Pt layer. We also characterize the spin Hall efficiencies of these two sets of samples using electrical ST-FMR measurements, and the results of which are shown in Table~\ref{tab:comparison}.

\section*{Acknowledgments}

The authors thank Daniel C. Ralph and Minh-Hai Nguyen for helpful discussions. This work was supported by AFOSR, under contract No. FA9550-14-1-0243. This work made use of the Cornell Center for Materials Research Shared Facilities which are supported through the NSF MRSEC program (DMR-1120296) as well as the Cornell NanoScale Facility, a member of the National Nanotechnology Coordinated Infrastructure (NNCI), which is supported by the National Science Foundation (Grant ECCS-15420819).

\section*{Author contributions}

G.D.F. and J.M.B. developed the measurement technique, F.G. and G.D.F. designed the experiments, F.G. and J.M.B. performed the experiments, F.G. the developed analysis method, F.G. and G.D.F. wrote the paper.

\section*{Competing financial interests}
The authors declare that they have no competing financial interests

\newpage

\pagebreak
\widetext
\begin{center}
\textbf{\large Supplemental Information for: \\ Ferromagnetic resonance phase imaging in spin Hall multilayers\\}
\vspace{5 mm}
\text{Feng Guo,$^1$ Jason M. Bartell,$^1$ and Gregory D. Fuchs$^1$}
\\
\textit{$^1$School of Applied and Engineering Physics, Cornell University, Ithaca, NY 14853, USA}
\vspace{10 mm}

\end{center}
\twocolumngrid

\setcounter{equation}{0}
\setcounter{figure}{0}
\setcounter{table}{0}
\setcounter{section}{0}
\setcounter{page}{1}
\makeatletter
\renewcommand{\theequation}{S\arabic{equation}}
\renewcommand{\thefigure}{S\arabic{figure}}
\renewcommand{\bibnumfmt}[1]{[S#1]}
\renewcommand{\thesection}{S\arabic{section}}
\section{Expansions of $\varphi_{\rm rf}^{\rm int}$ and $\varphi_{\textsc{\tiny FMR}}^{\rm int}$}

To gain more intuition, we expand $\theta^+$ and $\theta^-$ with respect to $h_{\rm ST}/h_{\rm Oe}^\parallel$ and $h_{\rm Oe}^\perp/h_{\rm Oe}^\parallel$, where $h_{\rm Oe}^\parallel$ and $h_{\rm Oe}^\perp$ are the in-plane and out-of-plane components of the Oersted field respectively. Thus we can rewrite Eq.~(\ref{eq:FMRphase}b) to:
\begin{subequations}
\label{eq:FMRphase_simple}
      \begin{align}
		\varphi_{\rm rf}^{\rm int}&=
		-\dfrac{\textstyle h_{\rm ST}}{\textstyle h_{\rm Oe}^\parallel}\nonumber \\
		&+\mathcal{O}\left[
			\left( \dfrac{\textstyle h_{\rm ST}}{\textstyle h_{\rm Oe}^\parallel} \right)
			\left( \dfrac{\textstyle h_{\rm Oe}^\perp}{\textstyle h_{\rm Oe}^\parallel} \right)^2
			+
			\dfrac{1}{3}\left( \dfrac{\textstyle h_{\rm ST}}{\textstyle h_{\rm Oe}^\parallel} \right)^3
			\right], \\
		\varphi_{\textsc{\tiny FMR}}^{\rm int}&=
		 {\textstyle h_{\rm Oe}^\perp \over \textstyle h_{\rm Oe}^\parallel} - 90^\circ \nonumber\\
		 &+\mathcal{O}\left[
			\left( \dfrac{\textstyle h_{\rm ST}}{\textstyle h_{\rm Oe}^\parallel} \right)^2
			\left( \dfrac{\textstyle h_{\rm Oe}^\perp}{\textstyle h_{\rm Oe}^\parallel} \right)
			+
			\dfrac{1}{3}\left( \dfrac{\textstyle h_{\rm Oe}^\perp}{\textstyle h_{\rm Oe}^\parallel} \right)^3
			\right].
	\end{align}		
\end{subequations}
As shown above, in the case of $h_{\rm ST}/h_{\rm Oe}^\parallel$, $h_{\rm Oe}^\perp/h_{\rm Oe}^\parallel \ll 1$ where the higher order terms become negligible, $\varphi_{\rm rf}^{\rm int}$ is primarily  sensitive to $h_{\rm ST}$ while $\varphi_{\textsc{\tiny FMR}}^{\rm int}$ is primarily sensitive to the Oersted field angle, as demonstrated in the main text.

\section{Phase references of current signal and FMR signal}

We raise a subtle point in measuring and defining the rf current phase. Although it is tempting to simply obtain the current phase from the current signal such as that in Fig.~\ref{fig:FMRphaseCenter} (c), there is a finite offset between the phase measured from the current signal [Fig.~\ref{fig:FMRphaseCenter} (c)] and that measured from the FMR signal [Fig.~\ref{fig:FMRphaseCenter} (d)]. To explain that, we need to account for the different time scales between FMR and current signals\cite{GuoBNF_prapplied15_TRANEtechnique}. The thermal gradient (corresponding to the FMR signal) has a faster response to the laser pulse and a quicker decay than these of the overall local temperature (corresponding to the current signal)\cite{BartellDLF_NatComm15_TRANE}. Therefore the difference in temporal profiles of the thermal gradient and the temperature creates an offset in the measured current phase.

\section{Derivation of ST-FMR signal with spatial varying FMR phase}

Here we include the spatial variation for both the FMR phase $\varphi_{\textsc{\tiny FMR}}(r)$ and amplitude $\theta_{\rm p}(r)$, while deriving the rectified voltage measured with ST-FMR. We use time dependent magnetoresistance and rf driving current to compute the rectified dc electric field as well as the mixing voltage, shown as Eq.~\ref{eq:Vmix} in the main text.



\begin{figure}[tb]
  \centering
  \includegraphics[width=0.65\figwidth]{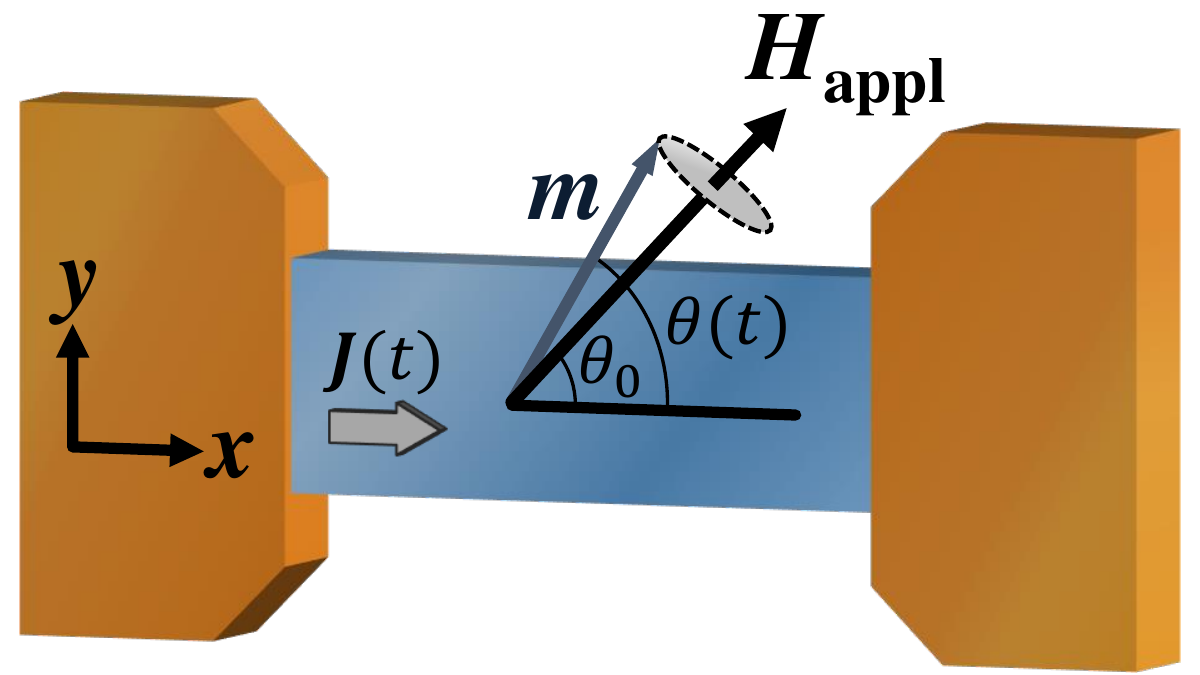}
  \caption{Diagram of ST-FMR measurement configuration. $\theta_0$ is the in-plane applied field angle with respect to $\hat{x}$ direction. $\theta(t)$ is the time varying magnetization orientation with respect to $\hat{x}$.
    \label{fig:SI_fig}}
\end{figure}

We start with the Ohm's law
\begin{align}
  \label{eq:1}
	{\bm E}&=\rho {\bm J},
\end{align}
and we assume the rf current density has a spatially uniform distribution and is flowing along $\hat{x}$: ${\bm J}(t)=J_0 \cos(\omega t+\varphi_{\rm rf})~\hat{x}$. To include anisotropic magnetoresistance (AMR) we write the sample's resistivity
\begin{align}
  \label{eq:2}
	\rho=\rho_0 + \Delta\rho\cos^2\theta(t),
\end{align}
where $\theta(t)=\theta_0+\theta_{\rm p}\cos(\omega t+\varphi_{\textsc{\tiny FMR}})$ is the angle between the current ($\hat{x}$) and magnetization vector, $\theta_{\rm p}$ the in-plane applied field angle and $\theta_{\rm p}$ is the precession amplitude, depicted in Fig.~\ref{fig:SI_fig}. We combine equations~\ref{eq:1} and \ref{eq:2} and apply trigonometric identities
\begin{align}
  \label{eq:3}
	E=	&\left\{ \rho_0+\Delta\rho\cos^2\left[ \theta_0+\theta_{\rm p} \cos(\omega t+\varphi_{\textsc{\tiny FMR}})\right]\right\}  \nonumber \\
		&\times J_0 \cos(\omega t+\varphi_{\rm rf}) \nonumber \\
	=&\bigg\{ \rho_0+\frac{1}{2}\Delta\rho\Big[ 1+\cos2\theta_0\cos\big[ 2\theta_{\rm p} \cos(\omega t +\varphi_{\textsc{\tiny FMR}}) \big]   \nonumber \\
	& -\sin2\theta_0 \sin\big[ 2\theta_{\rm p} \cos(\omega t +\varphi_{\textsc{\tiny FMR}}) \big]\Big]\bigg\} \nonumber \\
	&\times J_0 \cos(\omega t+\varphi_{\rm rf}).
\end{align}
For a small precession amplitude $\theta_{\rm p}\ll 1$, we can further simplify the Eq.~\ref{eq:3} using small angle approximations
\begin{align}
  \label{eq:4}
	E=	& \bigg\{ \rho_0+\frac{1}{2}\Delta\rho\Big[ 1+\cos2\theta_0  - 2\theta_{\rm p}\sin2\theta_0   \cos(\omega t +\varphi_{\textsc{\tiny FMR}}) \Big]\bigg\} \nonumber \\
	&\times J_0 \cos(\omega t+\varphi_{\rm rf}).
\end{align}
Since ST-FMR measures the mixed dc signal, we can drop all the oscillating terms such as $\cos(\omega t+\cdots)$ and $\cos(2\omega t+\cdots)$. The resulting dc electric field is 
\begin{align}
  \label{eq:5}
	{\bm E}_{\rm dc}=-\frac{1}{2} J_0 \theta_{\rm p}({\bm r}) \Delta\rho\sin2\theta_0  \cos\big[ \varphi_{\rm rf} - \varphi_{\textsc{\tiny FMR}}({\bm r}) \big] ~\hat{x}.
\end{align}

Next, we use the dc current to calculate the mixing voltage $V_{\rm}$ measured from ST-FMR
\begin{align}
  \label{eq:6}
	I_{\rm dc}=&\int \mathrm{d}y\, \mathrm{d}z  \frac{E_{\rm dc}}{\rho_0}, \\
	V_{\rm mix}=& I_{\rm dc} R_0 \nonumber \\
	=&- \frac{1}{2} J_0 \Delta R \sin2\theta_0 \int \mathrm{d}y\, \mathrm{d}z \theta_{\rm p}({\bm r})\, \cos\big[ \varphi_{\rm rf} - \varphi_{\textsc{\tiny FMR}}({\bm r}) \big],
\end{align}
where $\Delta R=R_0\, \Delta\rho/\rho_0$ is the resistance change due to the AMR effect. We only consider the spatial variation along y-direction, so we can rewrite the mixing voltage
\begin{align}
  \label{eq:8}
	V_{\rm mix}=-\frac{I_0 \Delta R  \sin2\theta_0}{2 w} \int \mathrm{d}y\,  \theta_{\rm p}(y)\, \cos\big[ \varphi_{\rm rf} - \varphi_{\textsc{\tiny FMR}}(y) \big],
\end{align}
where the rf current amplitude $I_0=J_0/(w t)$, $w$ is the channel width, and $t$ is the sample thickness.

In order to make connection to Eq.~\ref{eq:Vmix} in the main text, we include the field dependence of the phases. Note that we define $\varphi_{\textsc{\tiny FMR}}$ as the precession phase at the resonance field $H_{\rm app}=H_{\rm res}$. We substitute the FMR phase by $\varphi_{\textsc{\tiny FMR}} \to \varphi_{\textsc{\tiny FMR}} +\phi(H)$, where $\phi(H)$ is the field dependent precession phase. Thus we can rewrite Eq.~\ref{eq:8} as
\begin{align}
  \label{eq:9}
	V_{\rm mix}(H)=&-\frac{I_0 \Delta R  \sin2\theta_0}{2 w} \nonumber \\
		&\times \int \mathrm{d}y\,  \theta_{\rm p}(y)\, \cos\big[ \varphi_{\rm rf} - \varphi_{\textsc{\tiny FMR}}(y) - \phi(H) \big] \nonumber \\
		=&-\frac{I_0 \Delta R  \sin2\theta_0}{2 w} \nonumber \\
		& \times \Big\{ \chi'(H)\int \mathrm{d}y\,  \theta_{\rm p}(y)\cos[\varphi_{\rm rf} - \varphi_{\textsc{\tiny FMR}}(y)] \nonumber \\
		& +\chi''(H) \int \mathrm{d}y\, \theta_{\rm p}(y)\sin[\varphi_{\rm rf} - \varphi_{\textsc{\tiny FMR}}(y)] \Big\},
\end{align}
in which we use the relations $\chi'(H)=\cos\phi(H)$ and $\chi''(H)=\sin\phi(H)$. Eq.~\ref{eq:9} is essentially  Eq.~\ref{eq:Vmix} in the text.

For the uniform precession mode, which is typically assumed for ST-FMR analysis, the mixing voltage is reduced to
\begin{align}
  \label{eq:10}
	V_{\rm mix}^{\rm unif}(H)&=-\frac{I_0 \Delta R\,  \theta_{\rm p} \sin2\theta_0}{2} \Big\{ \chi'(H)  \cos(\varphi_{\rm rf}- \varphi_{\textsc{\tiny FMR}}) \nonumber \\
		&   +\chi''(H) \sin(\varphi_{\rm rf} - \varphi_{\textsc{\tiny FMR}}) \Big\}.
\end{align}
In the case of an in-plane driving field (i.e. in-plane Oersted field without spin torque), the FMR phase and the rf current phase have a simple relation $\varphi_{\rm rf} - \varphi_{\textsc{\tiny FMR}}=90^\circ$. Thus in the absence of spin torque, the $\chi'$ term vanishes and $V_{\rm mix}(H) \propto \chi''(H)$, corresponding to an antisymmetric spectral line shape. While the spin torque is present, the $\chi'$ becomes nonzero, and the spectrum is a linear combination of symmetric ($\chi'$) and antisymmetric ($\chi''$) components.

\section{Spatial variation of non spin Hall sample}

\begin{figure}[tb]
  \centering
  \includegraphics[width=\figwidth]{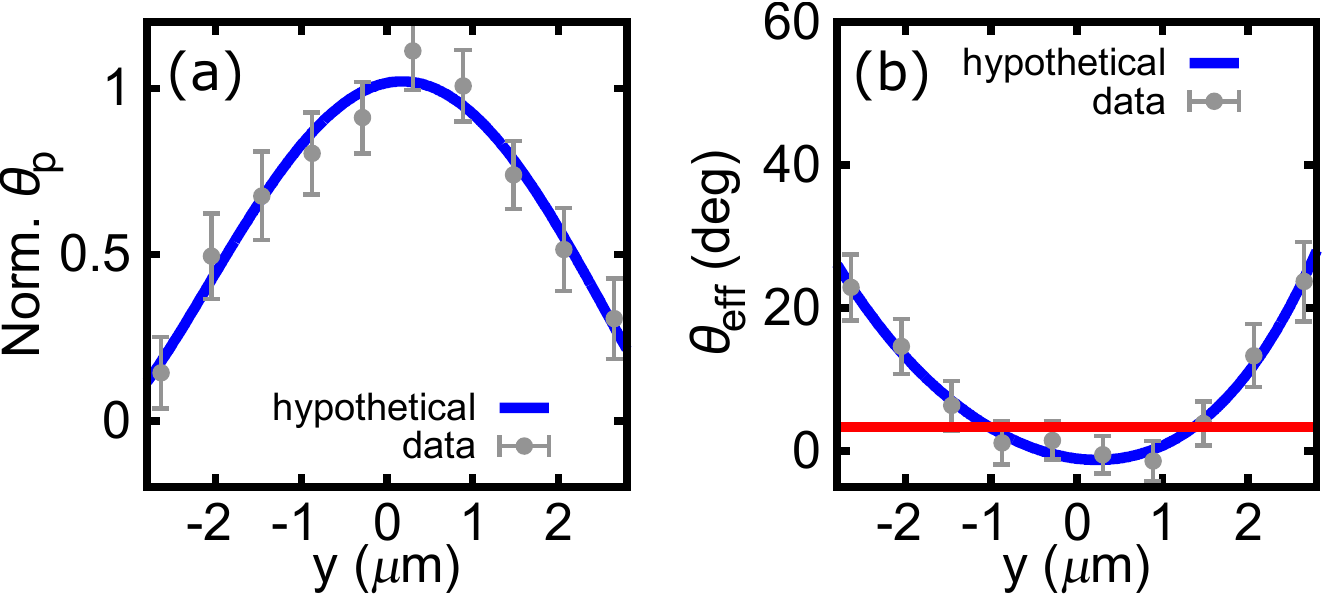}
  \caption{Spatial distributions for the non spin Hall sample with 2~nm Hf spacer: (a) the normalized precession amplitude and (b) effective driving field angle. The gray points in (a) and (b) are the data and the solid blue curves are the polynomial fits used for computing $\Delta$. The red line in (b) is the driving field angle, calculated from using Eq.~\ref{eq:st-fmr_phase}, that would be obtained from ST-FMR. The resulting phase correction $\Delta=(3.4\pm1.9)^\circ$.
    \label{fig:SI_fig}}
\end{figure}

Fig.~\ref{fig:SI_fig} presents the $y$ position dependent FMR amplitude and phase for the sample with 2~nm Hf spacer. Similar to the measurements in Fig.~\ref{fig:st-fmr_correction}, the observed FMR phase variation in the non spin Hall sample is nonuniform. The driving field titles about $25^\circ$ out of plane near the top and bottom edges. Despite the phase nonuniformity, the overall phase variation is smaller than that of the spin Hall sample. Consequently, the corresponding phase correction $\Delta=(3.4\pm1.9)^\circ$ is smaller than the spin Hall sample, as summarized in Table~\ref{tab:comparison}.

\end{document}